\documentstyle[prd,aps,epsf,multicol]{revtex}

\renewcommand{\narrowtext}{\begin{multicols}{2} \global\columnwidth20.5pc}
\renewcommand{\widetext}{\end{multicols} \global\columnwidth42.5pc}
\draft

\begin{document}

\bibliographystyle{prsty}

\title{Resummation of Nonalternating Divergent Perturbative Expansions}

\author{Ulrich D. Jentschura$^{1,2,}$\cite{InternetULJ}}

\address{$^1$National Institute of Standards and Technology,
Mail Stop 8401, Gaithersburg, MD 20899-8401, USA \\
$^2$Institut f\"{u}r Theoretische Physik, TU Dresden,
Mommsenstra\ss e 13, 01062 Dresden, Germany}

\maketitle

\begin{abstract}\noindent
A method for the resummation of nonalternating divergent perturbation
series is described. The procedure constitutes a generalization of the
Borel-Pad\'{e} method. Of crucial importance is a special integration
contour in the complex plane. Nonperturbative imaginary contributions
can be inferred from the purely real perturbative coefficients. A
connection is drawn from the quantum field theoretic problem of
resummation to divergent perturbative expansions in other areas of
physics.
\end{abstract}


\narrowtext 

%
%

In view of the probable divergence of quantum field theory in higher
order~\cite{Si1972,Li1977}, the resummation of the perturbation series
is necessary for obtaining finite answers to physical
problems. While the divergent expansions probably constitute
asymptotic series~\cite{We1999}, it is unclear whether unique answers can
be inferred from perturbation theory~\cite{Fi1997,Gr1994}. Significant
problems in the resummation are caused by infrared (IR) renormalons.
These are contributions corresponding to nonalternating divergent
perturbation series. The IR renormalons are responsible for the
Borel-nonsummability of a number of field theories including 
quantum chromodynamics (QCD) and quantum
electrodynamics (QED)~\cite{Fi1997,ItPaZu1977}.
 
%
%

\begin{figure}[htb]
\begin{center}
\begin{minipage}{8cm}
\caption{\label{contours} Integration contours for the evaluation of the
generalized Borel integral in Eq.~\protect{(\ref{method})}. }
\centerline{\mbox{\epsfysize=7.3cm\epsffile{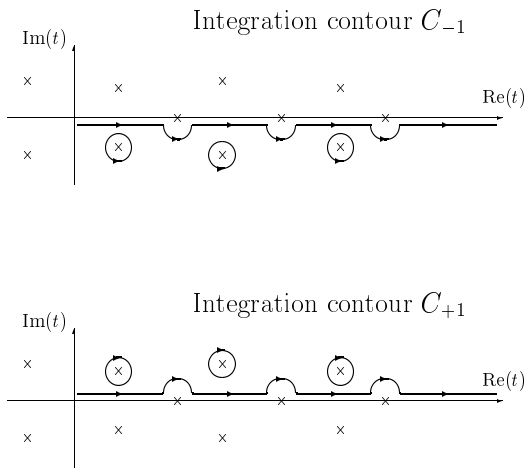}}}
\end{minipage}
\end{center}
\end{figure}

Here I advocate a modification of the resummation method proposed
in~\cite{Gr1994,Raczka1991} for nonalternating divergent perturbation
series. The method starts with a given input series,
\begin{equation}
\label{input}
f(g) \sim \sum_{n=0}^{\infty} c_n\,g^n \,, \qquad
c_n > 0 \,, \qquad g > 0\,,
\end{equation}
where $g$ is the coupling parameter and the perturbative coefficients
$c_n$ are expected to diverge as follows~\cite{VaZa1994},
\begin{equation}
\label{large}
c_n \sim K\,\frac{n!\,n^\gamma}{S^n} \,, \qquad
n \to \infty\,,
\end{equation}
with $K$, $\gamma$ and $S$ being constant.
The Borel transform $f_{\rm B}$ of the perturbation series (\ref{input})
\begin{equation}
\label{bt}
f_{\rm B}(g) = \sum_{n=0}^{\infty} 
  \frac{c_n}{n!}\,g^n
\end{equation}
has a finite radius of convergence about the origin. For the
evaluation of the Borel integral, $f_{\rm B}(g)$ has to be continued
analytically beyond the radius of convergence. Strictly speaking, this
analytic continuation has to be done on the branch cut in view of the
nonalternating character of the series~(\ref{input}). This requirement
can be relaxed slightly by performing the analytic continuation into
regions where $g$ acquires at least an infinitesimal imaginary part $g
\to g \pm {\rm i}\,\epsilon$.  In this case, the analytic continuation
can be achieved by evaluating Pad\'{e} approximants~\cite{Pi1999}.
The first $n+1$ terms of the Borel transformed series (\ref{bt}) can
be used to construct a diagonal or off-diagonal Pad\'{e}
approximant~(for the notation see~\cite{Ba1975,JeBeWeSo1999})
\begin{equation}
\label{pade}
{\cal P}_n(z) = \bigg[ [\mkern - 2.5 mu [n/2] \mkern - 2.5 mu ] \bigg/
[\mkern - 2.5 mu [(n+1)/2] \mkern - 2.5 mu ]
\bigg]_{f_{\rm B}}\!\!\!\left(z\right)\,,
\end{equation}
where $[\mkern - 2.5 mu [x] \mkern - 2.5 mu ]$ denotes the integral
part of $x$.  The resummation is accomplished by constructing the
sequence of transforms $\{ {\cal T} f_n(g) \}_{n=0}^{\infty}$ where
\begin{eqnarray}
\label{method}
{\cal T}f_n(g) &=& \int_{C_j} {\rm d}t\,\exp(-t)\,{\cal P}_n(g\,t) \,,
\end{eqnarray}
and the integration contour $C_j$ ($j=-1,0,+1$) is as shown in
Fig.~\ref{contours} (for $j=-1$ and $j=+1$). The result obtained along
$C_{-1}$ is the complex conjugate of the result along $C_{+1}$.  The
arithmetic mean of the results of the integrations along $C_{-1}$ and
$C_{+1}$ is associated with $C_0$.  Therefore, the result along $C_0$
is real rather than complex.  The limit of the sequence $\{ {\cal T}
f_n(g) \}_{n=0}^{\infty}$ (provided it exists),
\begin{equation}
\label{limit}
\lim_{n \to \infty} {\cal T} f_n(g) = f(g)\,,
\end{equation}
is a plausible complete nonperturbative result inferred from the
perturbative expansion~(\ref{input}).  Which of the contours $C_j$
($j=-1,0,+1$) is chosen, has to be decided on the basis of additional
considerations which do not follow from perturbation theory alone.

The zeros of the denominator polynomial of the Pad\'{e} approximant
[see Eq.~(\ref{pade})] correspond to the poles of the integrand in
Eq.~(\ref{method}).  Denote by $t$ the integration variable for the
evaluation of the generalized Borel integral in Eq.~(\ref{method}),
then the poles lie at $t = z_i$ (where the index $i$ numbers the
poles) along the positive real axis (${\rm Im} \, z_i = 0$) and in the
complex plane (${\rm Im} \, z_i \neq 0$).  The poles lying on the
positive real axis are treated as half-poles encircled in the
mathematically positive sense for $C_{-1}$ and as half-poles encircled
in the mathematically negative sense for $C_{+1}$.  The contour
$C_{-1}$ encircles all poles at $t = z_i$ in the lower right quadrant
of the complex plane (${\rm Re} \, z_i > 0$, ${\rm Im} \, z_i < 0$) in
the positive sense (see Fig.~\ref{contours}). The contribution of
these poles should be added to the final result.  The contour $C_{+1}$
is understood to encircle all poles in the upper right quadrant of the
complex plane in the mathematically negative sense.  In general, the
integrations along $C_{-1}$ and $C_{+1}$ lead to a nonvanishing
imaginary part in the final result for $f(g)$ [see Eq.~(\ref{limit})],
although all the perturbative coefficients $c_n$ are by assumption
real and positive [see Eq.~(\ref{input})].  It might be interesting to
note that, as with any complex integration, it is permissible to
deform the integration contours shown in~Fig.~\ref{contours} in accord
with the Cauchy Theorem as long as all pole contributions are properly
taken into account.

%
%

This paper represents a continuation of previous work on the
subject~\cite{Gr1994,Raczka1991,Pi1999}.  The resummation method
defined in Eqs.~(\ref{input})--(\ref{limit}) differs
from~\cite{Gr1994} in the combination of Borel and Pad\'{e} techniques
and, if compared to the remarkable investigations
in~\cite{Raczka1991,Pi1999} on the resummation of QCD perturbation
series, in the integration contour used for the evaluation of the
generalized Borel integral.  It is argued here that, when the Borel
transform (\ref{bt}) is analytically continued with Pad\'{e}
approximants (\ref{pade}), the contribution of poles lying off the
positive real axis has to be taken into account in order to obtain
consistent results in the resummation (see Fig.~\ref{contours}).
In~\cite{Raczka1991,Pi1999} it is argued that the Borel integral
should be evaluated by principal value.  It could appear that the
$C_0$ contour corresponds to the principal-value prescription.
However, this is not necessarily the case, if there are poles present
which lie off the positive real axis (i.e., at $t = z_i$ with ${\rm
Re}\,z_i > 0$, ${\rm Im}\,z_i \neq 0$).  The contribution of these
poles not only modifies the imaginary, but also the real part of
the final nonperturbative result.  Of course, when there are no poles
lying off the positive real axis, as it is the case for the problems
discussed in~\cite{Raczka1991,Pi1999}, then the principal-value
prescription used in~\cite{Raczka1991,Pi1999} is equivalent to the
$C_0$ contour.  Because the result obtained along $C_0$ is real, this
contour should be used whenever the existence of an imaginary part is
discouraged by physical reasons.

It is important to mention that the method presented here is not the
only prescription currently available for the resummation of divergent
perturbative expansions in quantum field theory.  For example, the
$\delta$ transformation (see Eq.~(4) in~\cite{JeBeWeSo1999}) is a very
useful method for the resummation of divergent perturbation series.
The $\delta$ transformation has a number of appealing mathematical
properties, including rapid and numerically stable convergence, and it
has been shown to yield consistent results in many cases, including
applications from quantum field theory~\cite{JeBeWeSo1999} and from
other areas of physics~\cite{We1996d}.  Because the $\delta$
transformation fulfills an accuracy-through-order relation (see
Eq.~(9) in~\cite{JeBeWeSo1999}), it can be used to predict
perturbative coefficients.  The $\delta$ transformation is primarily
useful for alternating series.  It fails, in general, in the
resummation of the nonalternating series discussed here. {\em The
$\delta$ transformation and the resummation method introduced here
complement each other}.

Three applications of the resummation method defined in
Eqs.~(\ref{input})--(\ref{limit}) are considered below: (i) the QED
effective action in the presence of a constant background electric
field, (ii) a mathematical model series which simulates the expected
large-order behavior of perturbative coefficients in quantum field
theory, (iii) the perturbation series for the energy shift of an
atomic level in a constant background electric field (including the
auto-ionization width).  The nonperturbative imaginary contributions
obtained along $C_{-1}$ and $C_{+1}$ find a natural physical
interpretation in all cases considered.

%
%

The QED effective action, or vacuum-to-vacuum amplitude, in the
presence of a constant background electric field has been treated
nonperturbatively in~\cite{Sc1951,ItZu1980}, and the result is
proportional to the integral
\begin{equation}
\label{exactSE}
S(g_{\rm E}) = 
- \!\! \int\limits_{0 - {\rm i}\,\epsilon}^{\infty - {\rm i}\,\epsilon} 
\frac{{\rm d}s}{s^2} \!
\left\{\cot s \! - \! \frac{1}{s} \! + \! \frac{s}{3} \right\}
\exp\left[-\frac{1}{\sqrt{g_{\rm E}}} \, s\right]\,,
\end{equation}
where $g_{\rm E}$ is a coupling parameter proportional to the square
of the electric field strength, $g_{\rm E} = e^2\,E^2/m_{\rm e}^4$.
Here, $m_{\rm e}$ is the electron mass, and $e$ is the elementary
charge. The natural unit system ($\hbar = c = 1$) is used.  The
imaginary part of $S(g_{\rm E})$ is proportional to the
electron-positron pair production amplitude per space-time interval
[there is of course also a muon-antimuon pair-production amplitude,
obtained by the imaginary part of (\ref{exactSE}) under the
replacement $m_{\rm e} \to m_{\mu}$, which is not discussed here].
$S(g_{\rm E})$ has the following asymptotic expansion in the coupling
parameter,
\begin{equation}
\label{asympSE}
S(g_{\rm E}) \sim 16 \, 
\left[ \sum_{n=0}^{\infty} 
\frac{4^n | {\cal B}_{2 n + 4}|}{(2 n + 4) \, (2 n + 3) \, (2 n + 2)} \,
g_{\rm E}^{n+1} \right]
\end{equation}
where ${\cal B}_{2 n + 4}$ is a Bernoulli number.  In view of the
asymptotics
\begin{equation}
\frac{4^n | {\cal B}_{2 n + 4}|}{n^3}
\sim \frac{\Gamma(2n+2)}{\pi^{2n+4}}\,
\left[1 + {\rm O}\left(\frac{1}{n}\right)\right]\,,\,\quad
n \to \infty\,,
\end{equation}
the perturbative coefficients, which are nonalternating in sign,
diverge factorially in absolute magnitude. The asymptotic
series~(\ref{asympSE}) for $S(g_{\rm E})$ is taken as the input series
for the resummation process [Eq.~(\ref{input})], and a sequence of
transforms ${\cal T} S_n(g_{\rm E})$ is evaluated using the
prescription~(\ref{method}).  The results have to be compared to the
exact nonperturbative expression~(\ref{exactSE}).  This is done in
Table~\ref{tableSE} for $g_{\rm E} = 0.05$. The partial sums of the
asymptotic series~(\ref{asympSE}) are listed in the second column.

%
%
\begin{center}
\begin{minipage}{8.0cm}
\begin{table}[htb]
\caption{\label{tableSE} Resummation of the asymptotic series for the
QED effective action~(\protect{\ref{asympSE}}) in a constant
background electric field for $g_{\rm E} = 0.05$.  Results in the
third column are obtained by the method indicated in
Eq.~(\ref{method}) along the integration contour $C_{-1}$.  The
partial sums in the second column are obtained from the asymptotic
series~(\ref{asympSE}).}
\begin{center}
\begin{minipage}{8.0cm}
\begin{tabular}{ccl}
\multicolumn{1}{c}{\rule[-3mm]{0mm}{8mm}{$n$}} &
\multicolumn{1}{c}{\rule[-3mm]{0mm}{8mm}{partial sum}} &
\multicolumn{1}{c}{\rule[-3mm]{0mm}{8mm}{${\cal T}S_n(g_{\rm E})$}} \\
\hline
2  & $0.001~146~032$ &  $0.001~144~848 - {\rm i}\,7.70 \times 10^{-17}$ \\
3  & $0.001~146~705$ &  $0.001~146~639 - {\rm i}\,8.22 \times 10^{-11}$ \\
4  & $0.001~146~951$ &  $0.001~147~113 - {\rm i}\,3.54 \times 10^{-8}$ \\
5  & $0.001~147~087$ &  $0.001~147~264 - {\rm i}\,1.93 \times 10^{-8}$ \\
6  & $0.001~147~195$ &  $0.001~147~173 - {\rm i}\,3.15 \times 10^{-7}$ \\
7  & $0.001~147~310$ &  $0.001~147~113 - {\rm i}\,2.58 \times 10^{-7}$ \\
8  & $0.001~147~469$ &  $0.001~147~162 - {\rm i}\,2.30 \times 10^{-7}$ \\
9  & $0.001~147~743$ &  $0.001~147~165 - {\rm i}\,2.63 \times 10^{-7}$ \\
10 & $0.001~148~327$ &  $0.001~147~144 - {\rm i}\,2.53 \times 10^{-7}$ \\
11 & $0.001~149~825$ &  $0.001~147~157 - {\rm i}\,2.46 \times 10^{-7}$ \\
12 & $0.001~154~375$ &  $0.001~147~155 - {\rm i}\,2.56 \times 10^{-7}$ \\
13 & $0.001~170~560$ &  $0.001~147~151 - {\rm i}\,2.51 \times 10^{-7}$ \\
14 & $0.001~237~137$ &  $0.001~147~156 - {\rm i}\,2.51 \times 10^{-7}$ \\
15 & $0.001~550~809$ &  $0.001~147~153 - {\rm i}\,2.53 \times 10^{-7}$ \\
16 & $0.003~228~880$ &  $0.001~147~154 - {\rm i}\,2.51 \times 10^{-7}$ \\
17 & $0.013~345~316$ &  $0.001~147~154 - {\rm i}\,2.52 \times 10^{-7}$ \\
18 & $0.081~610~937$ &  $0.001~147~153 - {\rm i}\,2.52 \times 10^{-7}$ \\
19 & $0.594~142~371$ &  $0.001~147~154 - {\rm i}\,2.52 \times 10^{-7}$ \\
20 & $4.852~426~276$ &  $0.001~147~154 - {\rm i}\,2.52 \times 10^{-7}$ \\
\hline
\multicolumn{1}{l}{exact} & 
  $0.001~147~154$ & $0.001~147~154 - {\rm i}\,2.52 \times 10^{-7}$ \\
\end{tabular}
\end{minipage}
\end{center}
\end{table}
\end{minipage}
\end{center}

Numerical results from perturbation theory are normally obtained by
(optimal) truncation of the perturbation series.  For the example
considered, (i) the partial sums do not account for the imaginary part
and (ii) due to the divergence of the perturbative expansion, no
improvement in the final result could be obtained by adding more than
the first 7 perturbative terms.  It requires a valid resummation
procedure to go beyond the accuracy obtainable by optimal truncation
of the perturbation series.  The transforms ${\cal T}S_n(g_{\rm E})$
displayed in the third column of Table~\ref{tableSE} apparently
converge to the full nonperturbative result given in
Eq.~(\ref{exactSE}), and the nonperturbative imaginary part, which
corresponds to the pair production amplitude, is reproduced although
the input series~(\ref{asympSE}) has purely real perturbative
coefficients.

%
%

Two specific mathematical model series are considered next.  The
series
\[
{\cal N}(g) \sim \sum_{n=0}^{\infty} n!\,g^n 
\]
has been used as a paradigmatic example for nonalternating divergent
series in the literature~\cite{We1999,ElGaKaSa1995,ElGaKaSa1996}.
This series can be resummed by the method~(\ref{method}). Moreover,
this resummation is even exact for all transformation orders $n \geq
2$.  This can be seen as follows.  The Borel transform
\[
{\cal N}_{\rm B}(g) = \sum_{n=0}^{\infty} g^n = 1/(1-g)
\]
is a geometric series. The summation of geometric series inside and
outside of the circle of convergence by Pad\'{e} approximants is exact
in all transformation orders $n \geq 2$.  So, for all $n \geq 2$ the
transforms ${\cal T}{\cal N}_n(g)$ fulfill the equality ${\cal T}{\cal
N}_n(g) = -1/g\,\exp(-1/g)\,\Gamma(0,-1/g) = {\cal N}(g)$, where
$\Gamma(0,x)$ is the incomplete Gamma function
(see~\cite{Ba1953vol1}), and the choice of the contour ($C_{-1}$ or
$C_{+1}$) determines on which side of the branch cut the incomplete
Gamma function is evaluated.

The asymptotic series,
\begin{equation}
\label{defM}
{\cal M}(g) \sim \sum_{n=0}^{\infty}
  \frac{\Gamma(n + \gamma)}{\Gamma(n)} \, n! \, g^n\,,
\end{equation}
constitutes a more interesting application of the resummation method
than ${\cal N}(g)$.  On account of the asymptotics,
\begin{equation}
\frac{\Gamma(n + \gamma)}{\Gamma(n)}
\sim n^\gamma \, \left( 1 + {\rm O}\left(\frac{1}{n}\right) \right)\,,
\quad n \to \infty\,,
\end{equation}
the series ${\cal M}(g)$ serves as a model for the expected
large-order behavior of perturbative coefficients in quantum field
theory [see Eq.~(\ref{large})].  The analytic summation of
(\ref{defM}) leads to
\begin{equation}
\label{exactM}
{\cal M}(g) = \Gamma(\gamma) \, 
\left(g\,\frac{\partial}{\partial g}\right)\,
{}_2 F_0(1,\gamma;g)\,, 
\end{equation}
where the hypergeometric ${}_2 F_0$ function has a branch cut along
the positive real axis (see~\cite{Ba1953vol1}).  The imaginary part
of~(\ref{exactM}) for $g > 0$ as a function of $g$ and $\gamma$ is
${\rm Im} \, {\cal M}(g) = \pi \, (1 - g\,\gamma) \, g^{-\gamma-1} \, 
\exp(-1/g)$,
where the integration is assumed to have been performed along the
contour $C_{+1}$.  For $C_{-1}$, the sign of the imaginary part is
reversed.  The numerical example considered here is $\gamma = 2.3$, $g
= 0.1$. In the Table~\ref{tableM}, numerical results are displayed for
the $n$th partial sums of the asymptotic series~(\ref{defM}) and the
transforms ${\cal T}{\cal M}_n(g)$ calculated according to
Eq.~(\ref{method}) in the range $n = 2,\dots,12$. While the partial
sums eventually diverge, the transforms ${\cal T}{\cal M}_n(g)$
exhibit apparent convergence to about 6 significant figures in ($n =
12$)th transformation order, and the transforms reproduce the
imaginary part although the coefficients of the series~(\ref{defM})
are all real rather than complex.  The integration is performed along
the contour $C_{+1}$.  The exact result in the last row of
Table~\ref{tableM} is obtained from Eq.~(\ref{exactM}). For the
evaluation of the transforms ${\cal T}{\cal M}_n(g)$ it is crucial to
use the contour $C_{+1}$ rather than a contour infinitesimally above
the real axis. For example, in order to obtain consistent numerical
results, it is necessary to take into account the poles at $t = 9.99
\pm {\rm i}\,0.578$ in ($n = 11$)th transformation order, 
encountered in the evaluation of the transform ${\cal T}{\cal
M}_{11}(g)$ according to Eq.~(\ref{method}), and the pole at $t = 9.99
\pm {\rm i}\, 0.495$ in ($n = 12$)th order for the evaluation of
${\cal T}{\cal M}_{12}(g)$.  These poles approximately correspond to
the triple pole at $t = 1/(0.1) = 10$ which would be expected in the
case $\gamma = 2$.

%
\begin{center}
\begin{minipage}{8.0cm}
\begin{table}[htb]
\caption{\label{tableM} Resummation of the model series
(\protect{\ref{defM}}) for $\gamma = 2.3$, $g = 0.1$
by the method indicated in Eq.~(\ref{method})
along the integration contour $C_{+1}$.
The partial sums are obtained from
the asymptotic series~(\ref{defM}).}
\begin{center}
\begin{minipage}{8.0cm}
\begin{tabular}{ccr}
\multicolumn{1}{c}{\rule[-3mm]{0mm}{8mm}{$n$}} &
\multicolumn{1}{c}{\rule[-3mm]{0mm}{8mm}{partial sum}} &
\multicolumn{1}{c}{\rule[-3mm]{0mm}{8mm}{${\cal T}{\cal M}_n(g)$}} \\
\hline
2 &  $0.445~451$ &  $0.393~554 + {\rm i}\,0.373~912$ \\
3 &  $0.559~685$ &  $0.840~561 + {\rm i}\,0.446~830$ \\
4 &  $0.640~410$ &  $0.764~942 + {\rm i}\,0.274~640$ \\
5 &  $0.703~981$ &  $0.765~339 + {\rm i}\,0.218~156$ \\
6 &  $0.759~669$ &  $0.763~012 + {\rm i}\,0.219~638$ \\
7 &  $0.813~594$ &  $0.762~186 + {\rm i}\,0.219~197$ \\
8 &  $0.870~909$ &  $0.762~196 + {\rm i}\,0.219~126$ \\
9 &  $0.937~322$ &  $0.762~224 + {\rm i}\,0.219~123$ \\
10 & $1.020~707$ &  $0.762~225 + {\rm i}\,0.219~127$ \\
11 & $1.133~528$ &  $0.762~223 + {\rm i}\,0.219~127$ \\
12 & $1.297~220$ &  $0.762~223 + {\rm i}\,0.219~127$ \\
\hline
\multicolumn{1}{l}{exact} & $0.762~223$ & $0.762~223 + {\rm i}\,0.219~127$ \\
\end{tabular}
\end{minipage}
\end{center}
\end{table}
\end{minipage}
\end{center}

%
%

When an atom is brought into an electric field, the levels become
unstable against auto-ionization, i.e.~the energy levels ${\cal E}$
acquire a width $\Gamma$ (${\cal E} \to {\rm Re} \, {\cal E} - {\rm
i}\Gamma/2$ where $\Gamma$ is the width). Perturbation theory cannot
account for the width. The coefficients are real, not
complex~\cite{Si1978}.  An established method for the determination of
the width is by numerical diagonalization of the Hamiltonian
matrix~\cite{HeInBr1974,DaKo1976,DaKo1978}. It is argued here that the
full complex energy eigenvalue, including the width, can also be
inferred from the divergent perturbation series by the resummation
method defined in Eqs.~(\ref{input})--(\ref{limit}), where the
appropriate integration contour is $C_{+1}$.  Perturbative
coefficients for the energy shift in arbitrarily high order can be
inferred from the Eqs.~(9,13--15,28--33,59--67,73) in~\cite{Si1978}.

The symmetry of the problem suggests the introduction of the parabolic
quantum numbers $n_1$, $n_2$ and $m$~\cite{LaLi1958} (the principal
quantum number is $n = n_1 + n_2 + m + 1$).  Here, calculations are
performed for the ground state with parabolic quantum numbers $n_1 =
0$, $n_2 = 0$, $m = 0$ and two L shell states, both of which are
coherent superpositions of the 2S and 2P states.  One of the L shell
states investigated here has the parabolic quantum numbers $n_1 = 1$,
$n_2 = 0$, $m = 0$, and the other L shell state has the quantum
numbers $n_1 = 0$, $n_2 = 1$, $m = 0$.  The Stark effect is
interesting because, depending on the atomic state, the perturbation
series are either completely nonalternating in sign (e.g., for the
ground state), or they constitute nonalternating divergent series with
a subleading divergent alternating component (e.g., for $n_1 = 0$,
$n_2 = 1$, $m = 0$), or the series are alternating with a subleading
divergent nonalternating component (e.g., for $n_1 = 1$, $n_2 = 0$, $m
= 0$).  The perturbation series for the Stark effect do not strictly
fulfill the assumptions of Eq.~(\ref{input}), and {\em the successful
resummation of these series might indicate that the method introduced
here is in fact more generally applicable}.  The large-order
asymptotics of the perturbative coefficients for the Stark effect are
given in Eqs.~(4,5) in~\cite{SiAdCiOt1979}.  In quantum field theory,
the alternating and nonalternating components correspond to
ultraviolet (UV) and IR renormalons.  Using the first 20 coefficients
of the perturbation series for the energy and evaluating the first 20
transforms according to Eq.~(\ref{method}), estimates for the real
part of the energy (Stark energy shift) and the imaginary part of the
energy (decay width of the state) may be obtained.  The apparent
convergence of the first 20 transforms for the real part of the energy
extends to 6--8 significant figures, whereas the convergence of the
imaginary part is much slower (2--3 significant figures).  In all
cases considered, both the real and the imaginary part of the energy
obtained by resummation compare favorably with values for the decay
width obtained by numerical diagonalization of the Hamiltonian
matrix~\cite{HeInBr1974,DaKo1976,DaKo1978}.  Here we concentrate on
the decay width, the full calculation will be described in detail
elsewhere.  The atomic unit system is used in the sequel, as is
customary for this type of
calculation~\cite{Si1978,HeInBr1974,DaKo1976,DaKo1978}.
%
%
In the atomic unit system, the unit
of energy is $\alpha^2\,m_{\rm e}\,c^2 = 
27.211\,{\rm eV}$ where $\alpha$ is the fine structure constant,
and the unit for the electric field is the field strength
felt by an electron at a distance of one Bohr radius $a_{\rm Bohr}$
to a nucleus of elementary charge, which is
$1/(4\,\pi\,\epsilon_0)\,(e/a_{\rm Bohr}^2) = 5.142\times 10^{11}\,
{\rm V}/{\rm m}$ (here, $\epsilon_0$ is the permittivity 
of the vacuum). 

Evaluations have been performed for all atomic levels and field
strengths of Table III in~\cite{SiAdCiOt1979}.  Three examples are
presented here.  For the ground state, at an electric field strength
of $E = 0.1$ in atomic units, the imaginary part of the first 20
transforms calculated according to Eq.~(\ref{method}) exhibits
apparent convergence to $\Gamma = 1.46(5) \times 10^{-2}$ which has to
be compared to $\Gamma = 1.45 \times 10^{-2}$ obtained from numerical
diagonalization of the Hamiltonian matrix~\cite{HeInBr1974}.  For the
L shell state with quantum numbers $n_1 = 0$, $n_2 = 1$, $m = 0$, at a
field strength of $E = 0.004$, the first 20 transforms exhibit
apparent convergence to an imaginary part of $\Gamma = 4.46(5) \times
10^{-6}$ which compares favorably to $\Gamma = 4.45 \times 10^{-6}$
from~\cite{DaKo1978}.  The most interesting case is the state $n_1 =
1$, $n_2 = 0$, $m = 0$, for which the nonalternating component of the
perturbation series is subleading.  At $E = 0.006$, resummation of the
complete perturbation series (including the leading alternating part)
leads to a decay width of $\Gamma = 6.08(5) \times 10^{-5}$, which is
again consistent with the result of $\Gamma = 6.09 \times 10^{-5}$
from~\cite{DaKo1978}.  The contour $C_{+1}$ is crucial, due to poles
lying off the real axis.

%
%
With the help of Carleman's Theorem~\cite{Ca1926} it is possible to
formulate a criterion which guarantees that there is a one-to-one
correspondence between a function and its associated asymptotic series
(see for example~\cite{GrGrSi1970}, Theorems XII.17 and XII.18 and the
definition on p.~43 in~\cite{ReSi1978}, p.~410 in~\cite{BeOr1978}, or
the comprehensive and elucidating 
review~\cite{Fi1997}).  Let $f (z)$ be a function
which is analytic in the interior and continuous on a sectorial region
${\cal S} = \{ z \vert \vert \arg (z) \vert \le k\,\pi/2 + \epsilon, 0
< \vert z \vert < R \}$ of the complex plane for some $\epsilon >
0$. Let the function $f$ have an asymptotic expansion $f (z) \; \sim
\; \sum_{n=0}^{\infty} \, c_n \, z^n$ (for $z \to 0$). The function
$f$ obeys a strong asymptotic condition (of order $k$) if there are
suitable positive constants $C$ and $\sigma$ such that $| f (z) \, -
\, \sum_{n=0}^m \, c_n \, z^n | \le C \, \sigma^{m+1} \, [k\,(m+1)]!
\, |z|^{m+1}$ holds for all $m$ and for all $z \in {\cal S}$. The
validity of such a condition implies that the function $f (z)$ is
uniquely determined by its asymptotic series (see Theorem XII.19 of
\cite{ReSi1978}).  Typically, series which entail nonperturbative
(imaginary) contributions do not fulfill the Carleman condition.  The
resulting ambiguity is reflected in the three integration contours in
Fig.~\ref{contours}, only one of which gives the physically correct
result. 

It has not escaped our attention that specialized variants of the method
introduced here can be constructed in those cases where additional
information about the perturbative coefficients (large-order
asymptotics, location of poles in the Borel plane, etc.) is available. 

%
%
Finite and consistent answers in quantum
field theory are obtained after regularization, renormalization and
resummation. Using a resummation method, as shown in the
Tables~\ref{tableSE} and~\ref{tableM}, it is possible to go beyond the
accuracy obtainable by optimal truncation of the perturbation series.
The purpose of resummation is to eventually reconstruct the full
nonperturbative result from the divergent perturbation series (see
also~\cite{Gr1994}).  I have examined two physical examples, the QED
effective action in a constant background electric field
[Eq.~(\ref{asympSE})] and the Stark energy shift.  The perturbation
series for the Stark effect contains nonalternating and alternating
divergent contributions, which correspond in their mathematical
structure to IR and UV renormalons in quantum field theory,
respectively.  It has been shown in each case that complete
nonperturbative results, including the pair-production amplitude for
electron-positron pairs and the atomic decay width, can be inferred
from the divergent nonalternating perturbation series by the
resummation method defined in Eqs.~(\ref{input})--(\ref{limit}).  A
mathematical model series (\ref{defM}), which simulates the expected
large-order growth of perturbative coefficients in quantum field
theory~[see Eq.~(\ref{large})], can also be resummed by the proposed
method (see Table~\ref{tableM}).  In all cases considered, the full
nonperturbative result involves an imaginary part, whereas the
perturbative coefficients are real.  The advocated method of
resummation makes use of the Pad\'{e} approximation applied to the
Borel transform of the divergent perturbation series. Advantage is
taken of the special integration contours $C_j$ ($j=-1,0,1$) shown in
Fig.~\ref{contours}. 
The author acknowledges helpful discussions with G. Soff,
P. J. Mohr and E. J. Weniger.

\end{multicols}

\begin{thebibliography}{10}

\vspace*{-1.8cm}

\bibitem[*]{InternetULJ}
Electronic address: ulj@nist.gov.

\bibitem{Si1972}
B. Simon, Phys. Rev. Lett. {\bf 28},  1145  (1972).

\bibitem{Li1977}
L.~N. Lipatov, Zh. \'{E}ksp. Teor. Fiz. {\bf 72},  411  (1977), [JETP {\bf 45},
  216 (1977)].

\bibitem{We1999}
G. West, Los Alamos preprint hep-ph/9911416.

\bibitem{Fi1997}
J. Fischer, Int. J. Mod. Phys. A {\bf 12},  3625  (1997).

\bibitem{Gr1994}
P.~A. Grunberg, Phys. Lett. B {\bf 325},  441  (1994).

\bibitem{ItPaZu1977}
C. Itzykson, G. Parisi, and J.~B. Zuber, Phys. Rev. D {\bf 16},  996  (1977).

\bibitem{Raczka1991}
P.~A. R\c{a}czka, Phys. Rev. D {\bf 43},  R9  (1991).

\bibitem{VaZa1994}
A.~I. Vainshtein and V.~I. Zakharov, Phys. Rev. Lett. {\bf 73},  1207  (1994).

\bibitem{Pi1999}
M. Pindor, Los Alamos preprint hep-th/9903151.

\bibitem{Ba1975}
G.~A. Baker, {\em Essentials of Pad\'{e} approximants} (Academic Press, New
  York, 1975).

\bibitem{JeBeWeSo1999}
U. Jentschura, J. Becher, E. Weniger, and G. Soff, Los Alamos preprint
  hep-ph/9911265.

\bibitem{We1996d}
E.~J. Weniger, Phys. Rev. Lett. {\bf 77},  2859  (1996).

\bibitem{Sc1951}
J. Schwinger, Phys. Rev. {\bf 82},  664  (1951).

\bibitem{ItZu1980}
C. Itzykson and J.~B. Zuber, {\em Quantum Field Theory} (McGraw-Hill, New York,
  NY, 1980).

\bibitem{ElGaKaSa1995}
J. Ellis, E. Gardi, M. Karliner, and M.~A. Samuel, Los Alamos preprint
  hep-ph/9509312.

\bibitem{ElGaKaSa1996}
J. Ellis, E. Gardi, M. Karliner, and M.~A. Samuel, Phys. Lett. B {\bf 366},
  268  (1996).

\bibitem{Ba1953vol1}
H. Bateman, {\em Higher Transcendental Functions} (McGraw-Hill, New York, NY,
  1953), Vol.~1.

\bibitem{Si1978}
H. Silverstone, Phys. Rev. A {\bf 18},  1853  (1978).

\bibitem{HeInBr1974}
M. Hehenberger, H.~V. McIntosh, and E. Br\"{a}ndas, Phys. Rev. A {\bf 10},
  1494  (1974).

\bibitem{DaKo1976}
R.~J. Damburg and V.~V. Kolosov, J. Phys. B {\bf 9},  3149  (1976).

\bibitem{DaKo1978}
R.~J. Damburg and V.~V. Kolosov, J. Phys. B {\bf 11},  1921  (1978).

\bibitem{LaLi1958}
L.~D. Landau and E.~M. Lifshitz, {\em Quantum Mechanics {\em (Volume 3 of the
  Course of Theoretical Physics)}} (Pergamon Press, London, 1958).

\bibitem{SiAdCiOt1979}
H. Silverstone, B.~G. Adams, J. $\check{\rm C}{\rm i}\check{\rm z}$ek, and P.
  Otto, Phys. Rev. Lett. {\bf 43},  1498  (1979).

\bibitem{Ca1926}
T. Carleman, {\em Les Fonctions Quasi-Analytiques} (Gauthiers-Villars, Paris,
  1926).

\bibitem{GrGrSi1970}
S. Graffi, V. Grecchi, and B. Simon, Phys. Lett. B {\bf 32},  631  (1970).

\bibitem{ReSi1978}
M. Reed and B. Simon, {\em Methods of Modern Mathematical Physics IV: Analysis
  of Operators} (Academic Press, New York, 1978).

\bibitem{BeOr1978}
C.~M. Bender and S.~A. Orszag, {\em Advanced Mathematical Methods for
  Scientists and Engineers} (McGraw-Hill, New York, NY, 1978).

\end{thebibliography}
\end{document}